\documentclass[12pt]{article}
\usepackage[english]{babel}
\usepackage{graphicx,epsfig}
\makeindex \textwidth 170mm \textheight 220mm \topmargin -5mm
\newcommand{\be}{\begin{equation}}
\newcommand{\ee}{\end{equation}}
\newcommand{\bea}{\begin{eqnarray}}
\newcommand{\eea}{\end{eqnarray}}

\def\rfr#1{eq. (\ref{#1})}
\def\rfrs#1#2{eqs. (\ref{#1})-(\ref{#2})}

\def\bb{\bibitem}
\def\eqi{\begin{equation}}
\def\eqf{\end{equation}}
\def\eqia{\begin{eqnarray}}
\def\eqfa{\end{eqnarray}}
\def\btab{\begin{tabular}}
\def\etab{\end{tabular}}
\def\bar{\begin{array}}
\def\ear{\end{array}}

\def\leti{Lense--Thirring}

\def\lb#1{\label{#1}}

\def\ic{inclination}

\begin{document}
\begin{titlepage}
\begin{flushright}
\today\\
BARI-TH/00\\
\end{flushright}
\vspace{.5cm}
\begin{center}
{\LARGE LAGEOS--type Satellites in Critical Supplementary Orbit
Configuration and the Lense--Thirring Effect Detection}
\vspace{1.0cm}
\quad\\
{Lorenzo Iorio$^{\dag}$\\ \vspace{0.1cm}
\quad\\
{\dag}Dipartimento di Fisica dell' Universit{\`{a}} di Bari, via
Amendola 173, 70126, Bari, Italy\\ \vspace{0.2cm} } \vspace{0.2cm}
\quad\\
{David M. Lucchesi$^{\sharp}$\\ \vspace{0.1cm}
\quad\\
{$\sharp$}Istituto di Fisica dello Spazio Interplanetario,
IFSI/CNR, Via Fosso del Cavaliere n. 100, 00133, Roma, Italy\\
\vspace{0.2cm} } \vspace{0.2cm}

{\bf Abstract\\}
\end{center}

{\noindent \small  In this paper we analyze quantitatively the
concept of LAGEOS--type satellites in critical supplementary orbit
configuration (CSOC) which has proven capable of yielding various
observables for many tests of General Relativity in the
terrestrial gravitational field, with particular emphasis on the
measurement of the Lense--Thirring effect. By using an entirely
new pair of LAGEOS--type satellites in identical, supplementary
orbits with, e.g., semimajor axes $a=12000$ km, eccentricity
$e=0.05$ and inclinations $i_{\rm S1}=63.4^{\circ}$  and $i_{\rm
S2}=116.6^{\circ}$, it would be possible to cancel out the impact
of the mismodelling of the static part of the gravitational field
of the Earth to a very high level of accuracy. The departures from
the ideal supplementary orbital configuration due to the orbital
injection errors would yield systematic gravitational errors of
the order of few percent, according to the covariance matrix of
the EGM96 gravity model up to degree $l=20$. However, the
forthcoming, new gravity models from the CHAMP and GRACE missions
should greatly improve the situation. So, it should be possible to
measure the gravitomagnetic shifts of the sum of their nodes
$\Sigma\dot\Omega^{\rm LT}$ with an accuracy level perhaps less
than 1$\%$, of the difference of their perigees
$\Delta\dot\omega^{\rm LT}$ with an accuracy level of 5$\%$ and of
$\dot X^{\rm LT}\equiv\Sigma\dot\Omega^{\rm
LT}-\Delta\dot\omega^{\rm LT}$ with an accuracy level of 2.8$\%$.
Such results, which are due to the non--gravitational
perturbations mismodelling, have been obtained for an
observational time span of about 6 years and could be further
improved by fitting and removing from the analyzed time series the
major time--varying perturbations which have known periodicities.
 }
\end{titlepage} \newpage \pagestyle{myheadings} \setcounter{page}{1}
\vspace{0.2cm} \baselineskip 14pt

\setcounter{footnote}{0}
\setlength{\baselineskip}{1.5\baselineskip}
\renewcommand{\theequation}{\mbox{$\arabic{equation}$}}
\noindent

\section{Introduction}
The idea of using a pair of twin satellites, denoted as S1 and S2,
in identical orbits with the same semimajor axes $a$ and
eccentricities $e$, except for the inclinations $i$ of their
orbital planes, which should be supplementary, in order to measure
the general relativistic Lense--Thirring effect (Lense and
Thirring 1918) in the gravitational field of the Earth\footnote{In
(Ciufolini {\it et al} 1998) an experimental check of such
prediction of General Relativity in the field of the Earth by
using the laser data of LAGEOS and LAGEOS II satellites is
reported. The claimed accuracy is of the order of 20$\%$.} was put
forth for the first time by Ciufolni with the proposed
LAGEOS--LAGEOS III mission (Ciufolini 1986). The proposed
observable is the sum of the rates of the longitudes of the
ascending nodes \eqi\Sigma\dot\Omega\equiv\dot\Omega_{\rm
S1}+\dot\Omega_{\rm S2}.\lb{sumn}\eqf Indeed, it turns out that
while the Lense--Thirring secular nodal rates are independent of
the inclinations of the satellites and add up in \rfr{sumn}, the
classical secular nodal rates induced by the oblateness of the
Earth, which would mask the relativistic effect due to the
uncertainties in the even zonal coefficients $\delta J_2,\ \delta
J_4,\ \delta J_6,...$ of the multipolar expansion of the
terrestrial gravitational field, are equal and opposite for
supplementary orbital planes because they depend on odd powers of
$\cos i$, so that they would be cancelled out by \rfr{sumn}. Later
on, the orbital and physical configuration of LAGEOS III slightly
changed: the eccentricity of its orbit was increased in order to
be able to perform other general relativistic tests, its mass was
reduced so to reduce the mission--launch costs, and the area was
reduced in such a way to guarantee the same area--to--mass ratio
of the older LAGEOS, so to reduce the impact of the
non--gravitational perturbations. Thus LARES was born (Ciufolini
1998). In Table 1 we quote the orbital parameters of some existing
or proposed laser--ranged satellites which are used, or could be
used, in general relativistic tests. The accuracy available with
the originally proposed version of the LAGEOS--LARES mission
should amount to 2$\%$--3$\%$ (Ciufolini  1998).
\begin{table}[ht!]
\caption{Orbital parameters of LAGEOS, LAGEOS II, LARES, S1 and
S2.} \label{para}
\begin{center}
\begin{tabular}{lllllll}
\noalign{\hrule height 1.5pt} Orbital parameter & LAGEOS & LAGEOS
II & LARES & S1 & S2\\ \hline
$a$ (km) & 12270 & 12163 & 12270 & 12000 & 12000\\
$e$ & 0.0045 & 0.014 & 0.04 & 0.05 & 0.05\\
$i$ (deg) & 110 & 52.65 & 70 & 63.4 & 116.6\\
\noalign{\hrule height 1.5pt}
\end{tabular}
\end{center}
\end{table}
Very recently, some modifications of the observable to be adopted
in the LARES mission have been suggested (Iorio {\it et al} 2002).
The total error should then become $\sim\ 1\%$.

The concept of satellites in identical and supplementary orbits
have been recently extended also to the perigees (Iorio 2002;
2003). In particular, it has been noticed that also the difference
of the rates of the perigees
\eqi\Delta\dot\omega\equiv\dot\omega_{\rm S1}-\dot\omega_{\rm
S2}\lb{diffp}\eqf could be considered, in principle, for measuring
the gravitomagnetic field of the Earth. Indeed, the
Lense--Thirring secular apsidal rates depend on $\cos i$ and add
up in \rfr{diffp}, while the classical secular apsidal rates due
to the oblateness of the Earth, which depend on $\cos^2 i$ and on
even powers of $\sin i$, are equal and cancel out in \rfr{diffp}.
Of course, such an observable could not be considered for the
LAGEOS--LARES mission since the eccentricity of LAGEOS is too
small and the perigee of its orbit is badly defined. On the
contrary, launching an entirely new pair of LAGEOS--type
satellites in rather eccentric orbits would allow to adopt both
\rfr{sumn} and \rfr{diffp} and also \eqi\dot
X\equiv\Sigma\dot\Omega-\Delta\dot\omega.\lb{sumdiff}\eqf In
(Iorio 2002) it has been noticed that it should be better to adopt
the critical inclinations $i_{\rm S1}=63.4^{\circ}$ and $i_{\rm
\rm S2}=116.6^{\circ}$ because, in this way, the periods of many
time--dependent
 harmonic orbital perturbations of gravitational and
 non--gravitational origin would be not too long. So, it would be
 possible to adopt an observational time span $T_{\rm obs}$ of a few
 years. This fact would be important not only from the point of
 view of reducing the data analysis time, but also because
 certain relevant and very useful assumptions on the surface properties of the
 satellites and on their spins motion, which would affect certain
 subtle but important non--gravitational
 perturbations,
 could be safely done by adopting just the first years of life of both
 satellites for the data analysis.

In this paper we wish to analyze quantitatively the impact of many
systematic errors induced by gravitational and non--gravitational
perturbations on the proposed observables so to yield realistic
error budgets for such new proposed gravitomagnetic experiments
and clarify if the alternative proposed observables are really
competitive with the sum of the nodes.

The paper is organized as follows. In section 2 we will deal with
the systematic error due to the mismodelling in the even zonal
harmonics of geopotential and its sensitivity to the orbital
injection errors in the inclinations of the satellites. In section
3 we will focus our attention on the impact of the
non--gravitational perturbations. Section 4 is devoted to the
conclusions.
\section{The gravitational errors}
Contrary to the originally proposed LAGEOS--LARES mission (Iorio
{\it et al}, 2002), with the new satellites S1 and S2 of Table 1,
which have $a_{\rm S1}=a_{\rm S2},\ e_{\rm S1}=e_{\rm S2}$, it
would be possible to cancel out exactly from \rfrs{sumn}{sumdiff}
the systematic errors due to the even zonal harmonics of the
geopotential, provided that the inclinations of the two orbital
planes are exactly supplementary.

Of course, this could not occur in reality because of the
unavoidable orbital injection errors in the Keplerian orbital
elements of the satellites, in particular those in the
inclinations. In Figure 1--Figure 3 we show the impact of the
deviations of the inclinations from the nominal condition of
supplementarity on the systematic errors in $\Sigma\dot\Omega,\
\Delta\dot\omega$ and $\dot X$.
\begin{figure}[ht!]
\begin{center}
\includegraphics*[width=13cm,height=10cm]{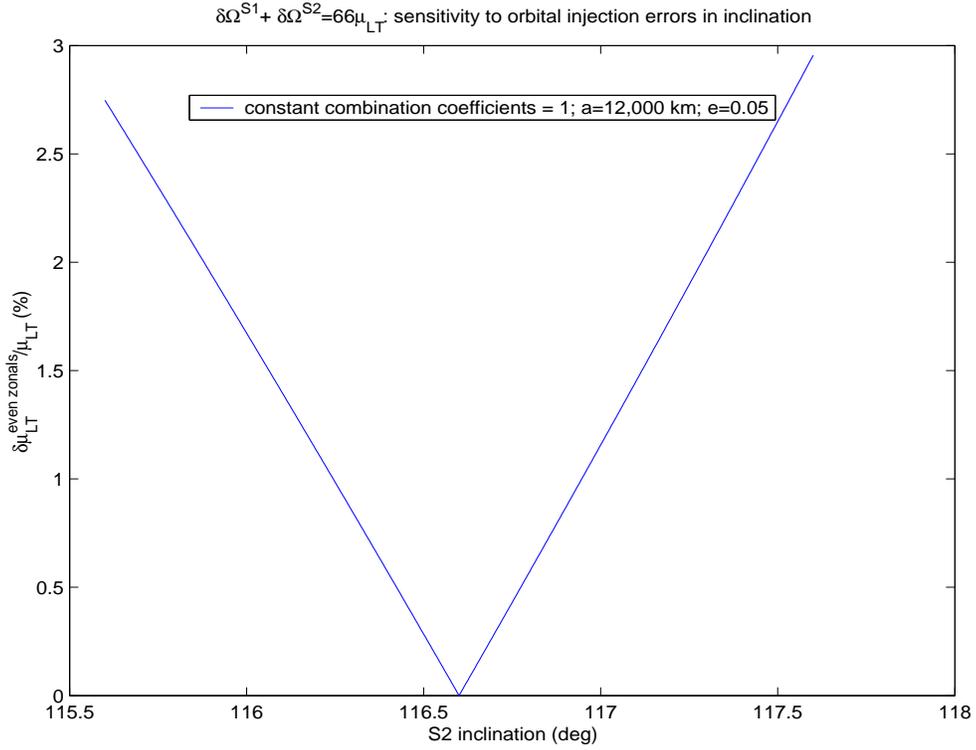}
\end{center}
\caption{\footnotesize Influence of the injection errors in the
\ic\ on the zonal  error of $\Sigma\dot\Omega$.} \label{figura1}
\end{figure}
\begin{figure}[ht!]
\begin{center}
\includegraphics*[width=13cm,height=10cm]{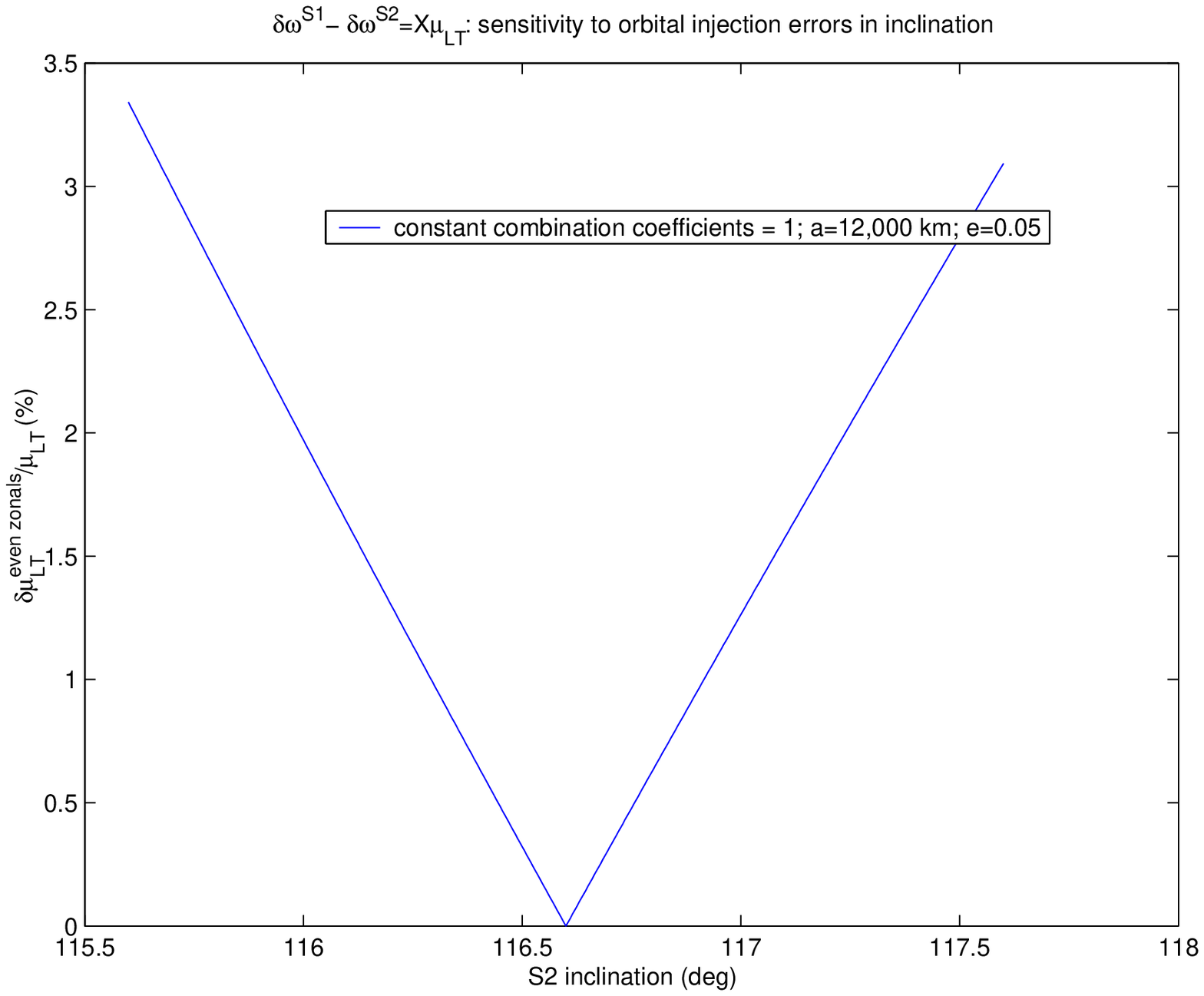}
\end{center}
\caption{\footnotesize Influence of the injection errors in the
\ic\ on the zonal  error of the $\Delta\dot\omega$.}
\label{figura2}
\end{figure}
\begin{figure}[ht!]
\begin{center}
\includegraphics*[width=13cm,height=10cm]{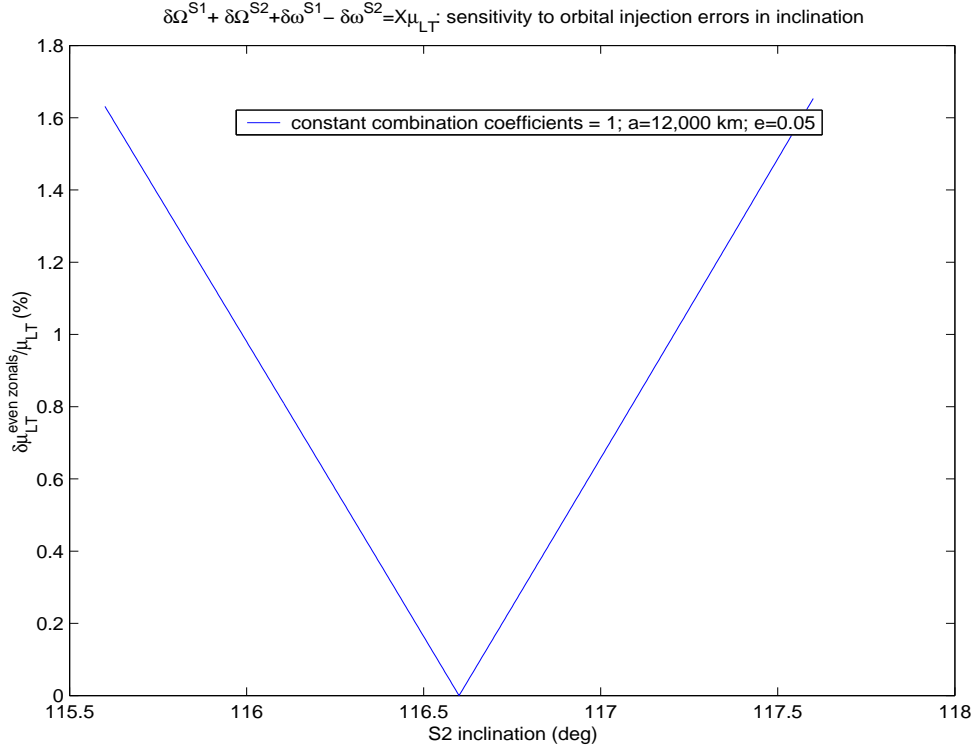}
\end{center}
\caption{\footnotesize Influence of the injection errors in the
\ic\ on the zonal  error of $\dot X$.} \label{figura3}
\end{figure}
The covariance matrix of the even zonal harmonics of the EGM96
Earth gravity model (Lemoine {\it et al}, 1998) up to degree
$l=20$ has been employed\footnote{However, it should be mentioned
that, according to (Ries {\it et al}, 1998), it would be
unjustified to extend the validity of the covariance matrix of
EGM96 to any generic time span of a few years due to secular,
seasonal and stochastic variations of the terrestrial
gravitational field which have not been included in the solution
of EGM96. Then, the validity of its covariance matrix should be
limited just to the multi--decadal time span during which the data
for its construction have been collected.}. It is worth noticing
that the results obtained here can be considered reliable even if
the higher degree terms of EGM96 would not be accurate enough
because the orbits of the LAGEOS--type satellites are almost
insensitive to the harmonics of degree higher than $l=20$. As it
can be noticed, for deviations of the order of 1
degree\footnote{In general, the orbital injection errors depend
strongly upon the final stage of the rocket. Indeed, if it uses a
solid propellant a conservative estimate yields $\delta i_{\rm
inj}\sim 1^{\circ}$, while with a liquid propellant, which is more
expensive, $\delta i_{\rm inj}\sim 0.6^{\circ}$ (L Anselmo,
private communication, 2002). For the LAGEOS III project the
predicted orbital injection error in the inclination, by using a
McDonnel--Douglas launcher, was estimated to be 0.03$^{\circ}$ at
3-$\sigma$ (See the McDonnell-Douglas document VDD5318 M2V, March
1987, cited in (Ciufolini, 1989)). } from the nominal
supplementary values, the systematic error due to the geopotential
would amount to few percent. However, it is really important to
notice that such estimates should be dramatically improved by the
new, more accurate terrestrial gravity models from CHAMP and GRACE
missions. Even without claiming any precise predictions which
could turn out to be too optimistic when the real data will be
finally available, it should be realistic to expect that the
systematic error due to the geopotential will fall well below the
1$\%$ level for all the observables considered here. It turns out
that for deviations of a few kilometers of the semimajor axes from
their nominal values the error due to the mismodelled even zonal
harmonics of the geopotential is well below 1$\%$ for the three
proposed observables.

In regard to the time--dependent gravitational perturbations of
tidal origin (Iorio 2001), the  perigees would be particularly
sensitive to them. However, as pointed out in (Iorio 2002), on one
hand, some very insidious tidal perturbations like that induced by
the 18-6 year constituent are cancelled out by the difference of
the perigees and, on the other, the choice of the critical
inclination for the two satellites would allow to reduce greatly
the periods of certain uncancelled perturbations which, then,
could be fitted and removed over a $T_{\rm obs}$ of just a few
years during which they would be able to describe some full
cycles. It turns out that also the time-dependent perturbations
induced on the perigee by the $J_{2n+1}$ odd zonal harmonics of
the geoptential cancel out in the difference of the perigees.
Moreover, it should be considered that also the uncertainty in the
time-varying part of the terrestrial gravitational field would be
reduced by the forthcoming models from CHAMP and GRACE, as can be
found in ${\rm http://op.gfz-de/grace/}$.
\section{The non--gravitational perturbations}
In view of the forthcoming improvements in the accuracy of our
knowledge of the terrestrial gravitational field, a major role in
the error budget of the proposed experiments will be played by the
perturbations due to the non--conservative forces.
Indeed,  concerning the measurement of the Lense-Thirring effect
using the existing LAGEOS satellites, we know that, according to
(Lucchesi 2001; 2002a), a crucial role in the definition of the
error budget is played by the mismodelling in the
non--gravitational perturbations (NGP). The perturbative effects
due to the visible radiation effects, as well as those related
with the thermal thrust perturbations---if not correctly
modelled---may cause errors in the observable orbital elements
comparable to or even larger than the relativistic secular shift.
We computed, through a numerical simulation and analysis
(considering the eclipses passages), the perturbative effects due
to these non--conservative forces on the proposed combinations.
Then, using the uncertainties of each perturbative model adopted
we estimated, in a conservative way, the error budget due to these
perturbations in the perigee and nodal rates of the proposed
satellites and in each of their combinations. We analyzed the
perturbative effects due to
\begin{itemize}
  \item direct visible solar radiation,
  \item Earth albedo radiation,
  \item terrestrial Yarkovsky--Rubincam radiation,
  \item solar Yarkovsky--Schach radiation,
  \item possible asymmetric reflectivity
\end{itemize}
Only the terrestrial Yarkovsky--Rubincam thermal thrust
perturbation produce secular effects in the satellites node and
perigee. Nevertheless these secular effects are usually
small--about 1 mas yr$^{-1}$ or less--and negligible when compared
to the long--term effects produced in the same elements by the
other non--gravitational perturbations (Lucchesi 2002a). Then, in
order to minimize the impact of these long-term effects, we
estimated an {\it Ideal Period} (IP) during which a large fraction
of the analyzed perturbations averages out. For this IP, tuned by
the larger perturbative effects, we obtained a value close to 6
years (2187 days). This IP has been estimated from the longest
periodicity, in the satellites perigee, due to direct solar
radiation pressure---the largest perturbative
effect---corresponding to the spectral line $\dot{\Omega}
+\dot{\lambda}-\dot{\omega}$ (729--day) of the S1 satellite. The
period obtained is a multiple integer of the other characteristics
periodicities of the analyzed NGP, and is very close to the period
of 734-day due to the node precession of the CSOC satellites, see
sub--section 3.1. This IP of about 6--year represents a good
compromise in the averaging out process of the non--gravitational
perturbations and, at the same time, on the accumulation of the
integrated orbital residuals over a time long enough to detect a
so tiny effect as the Lense--Thirring dragging. Anyway, for a
first measurement of the relativistic effect with the proposed
CSOC satellites, a minimum IP of about 2 years (729 days) may be
used. Of course, for a fixed IP, the averaging of the NGP depends
also from the initial conditions of the satellites configuration,
e.g., from the orientation of their orbits with respect to the
Sun. This aspect has been tested performing several simulations
over our 2187 days IP, but with different values for the initial
conditions of the satellites orbit in space---in the ascending
node and perigee initial positions---and for different periods of
the year, i.e., of the Earth position along its orbit with respect
to the Sun. In the following we present the results we obtained
for an initial configuration (the same for the two proposed
satellites) very close to that of LAGEOS when launched in May
1976. The results we obtained for the NGP mismodelling impact on
the suggested combinations of the satellites node and perigee, may
be considered a conservative approach to the final error budget.
Indeed, we have been able to obtain a better averaging of the NGP
effects using different configurations, but we also obtained
larger effects in the perigee and node rates for other initial
conditions. At the same time the proposed configuration for the
two satellites has the advantage to be realized without particular
difficulties using the actual launchers.

In regard to the Asymmetric Reflectivity effect we have made the
assumption that the asymmetry in the reflection from the
satellites surface is the same for the two satellites, and with
the same value as the one estimated in the case of LAGEOS.
Recently, in the case of LAGEOS II (Lucchesi 2002b), has been
estimated that a large fraction of this asymmetry in the albedoes
of the two hemisphere, could be explained by the specular
reflection of the visible solar--radiation from the Germanium
Cube--Corner--Retroreflectors (CCR) embedded in the satellite
surface. If this would be correct, we can probably neglect the
effect by not including these CCR on the satellite surface. But
the problem of a complete physical explanation of the anisotropy
in the case of LAGEOS satellites is still open, and possible other
mechanisms may contribute to the asymmetry in addition to the
Germanium CCR specular reflection of sun light. We then estimated
the effects of this perturbation using the empirical values
obtained in the case of LAGEOS and LAGEOS II (Lucchesi, 2002a). We
can look again to our final error budget as a conservative
estimate of the non-gravitational perturbative effects
uncertainties to the proposed derivation of the Lense--Thirring
precession on the combination of perigee and node of the CSOC
satellites.

The rest of the section is organized as follows. In sub--section
3.1 we shall describe some of the details of the analytical
results on the combinations of the proposed orbital elements of
the two CSOC satellites. In sub--section 3.2 the results of a
numerical simulation and analysis are given and briefly compared
with the analytical results.
\subsection{Analytical results}
The relativistic Lense--Thirring precessions on the node and the
perigee of the CSOC satellites are
\begin{equation}
\dot{\Omega}^{\rm LT}={2G\over {c^2a^3}}{J_{\oplus}\over
{(1-e^2)^{3/2}}}=32.9 \quad {\rm mas\ yr}^{-1} \label{OLT}
\end{equation}
for both satellites, and
\begin{equation}
\dot{\omega}^{\rm LT}=-{6G\over {c^2a^3}}{J_{\oplus}\over
{(1-e^2)^{3/2}}}\cos i = \left\{ \begin{array}{ll}
-44.2 \quad {\rm mas\ yr}^{-1} & \textrm{in the case of S1}\\
+44.2 \quad {\rm mas\ yr}^{-1} & \textrm{in the case of S2}
\end{array} \right. \label{omeLT}
\end{equation}
where $G$ is the Newtonian gravitational constant, $c$ is the
speed of light in vacuum and $J_{\oplus}$ is the Earth's angular
momentum. Then, for the combinations proposed in this work we
obtain
\begin{equation}
\Sigma\dot{\Omega}^{\rm LT}\equiv\dot{\Omega}^{\rm LT}_{\rm
S1}+\dot{\Omega}^{\rm LT}_{\rm S2}=65.8 \quad {\rm mas\ yr}^{-1}
\label{dueOLT}
\end{equation}
\begin{equation}
\Delta\dot{\omega}^{\rm LT}\equiv\dot{\omega}^{\rm LT}_{\rm
S1}-\dot{\omega}^{\rm LT}_{\rm S2}=-88.4 \quad {\rm mas\ yr}^{-1}
\label{diffome}
\end{equation}
\begin{equation}
\dot{X}^{\rm LT}\equiv\Sigma\dot{\Omega}^{\rm
LT}-\Delta\dot{\omega}^{\rm LT}=154.2 \quad {\rm mas\ yr}^{-1}
\label{combi}
\end{equation}
We will focus on the perigee rate of the proposed satellites
because of the larger perturbations and less accurate
determination of this element with respect to the node. The effect
of the direct solar radiation pressure on the perigee rate for a
spherically shaped, passive, laser--ranged satellite of
LAGEOS--type is
\begin{equation}
\dot{\omega}={3a_{\odot}\over {8nae}}\left\{ \begin{array}{lllll}
\Big [1-\Big(1+\cos i\Big)\cos \epsilon+\cos i\Big ]
\cos \Big(\Omega+\lambda+\omega \Big) + \\
\Big [1-\Big(1-\cos i\Big)\cos \epsilon-\cos i\Big ]
\cos \Big(\Omega+\lambda-\omega \Big) + \\
\Big [1+\Big(1+\cos i\Big)\cos \epsilon+\cos i\Big ]
\cos \Big(\Omega-\lambda+\omega \Big) + \\
\Big [1-\Big(1+\cos i\Big)\cos \epsilon-\cos i\Big ]
\cos \Big(\Omega-\lambda-\omega \Big) + \\
2\sin i \sin \epsilon \cos\Big(\lambda -\omega \Big)-2\sin i
\sin\epsilon \cos \Big(\lambda +\omega \Big),\\
\end{array}\right.
\label{per1}
\end{equation}
where $a_{\odot}$ is the acceleration due to direct solar
radiation pressure (about $3.6\times 10^{-9}$ m s$^{-2}$, as for
the LAGEOS satellites), $n$ is the satellite mean motion,
$\epsilon$ is the obliquity of the ecliptic and $\lambda$ the
Earth ecliptic longitude around the Sun. The periodicities of this
perturbations are also characteristic of the Earth radiation
pressure, i.e., the albedo effect, and of the solar
Yarkovsky--Schach thermal thrust effect. These are among the
largest perturbations on a satellite orbiting the Earth. Scharroo
et al. (1991) proved that also an Asymmetric Reflectivity of the
satellite hemispheres could be responsible of unmodelled effects
on LAGEOS semimajor axis. Successively, Metris et al. (1997)  and
Lucchesi (2002a) considered the effects of this perturbation on
LAGEOS and LAGEOS II eccentricity vector excitations and perigee
rate. This perturbation (Lucchesi 2002a) is characterized by
additional spectral lines with respect from those obtained with
Eq. (\ref{per1}). In Table 2 we report the spectral lines and the
periods of the main periodic contributions to the perigee rate
from the above cited NGP.
%
\begin{table}[ht!]
\caption{NGP spectral lines on the perigee rate of the proposed
satellites and their periodicities (days).} \label{para}
\begin{center}
\begin{tabular}{lllllll}
\noalign{\hrule height 1.5pt} Spectral line & S1 & S2\\ \hline
$\Omega+\lambda+\omega$  & 725.6 & 243.7 \\
$\Omega+\lambda-\omega$ & 728.6 & 244.0 \\
$\Omega-\lambda+\omega$ & 244.0 & 728.6 \\
$\Omega-\lambda-\omega$ & 243.7 & 725.6 \\
$\Omega+2\lambda+\omega$ & 243.0 & 146.2 \\
$\Omega+2\lambda-\omega$ & 243.3 & 146.3 \\
$\Omega-2\lambda+\omega$ & 146.3 & 243.3 \\
$\Omega-2\lambda-\omega$ & 146.2 & 243.0 \\
$\Omega+\omega$ & 735.4& 732.4 \\
$\Omega-\omega$ & 732.4 & 735.4 \\
$\lambda+\omega$ & 364.9 & 364.9 \\
$\lambda-\omega$ & 365.6 & 365.6 \\
$2\lambda+\omega$ & 182.5 & 182.5 \\
$2\lambda-\omega$ & 182.7 & 182.7 \\
$\omega$ & 365,351 & 365,351  \\
\noalign{\hrule height 1.5pt}
\end{tabular}
\end{center}
\end{table}
%
As we can see, our IP of about 2187 days is very close to an
integer multiple of the shorter--period lines obtained.
Eq. (\ref{per2}) gives the effect of direct solar radiation
pressure on the perigee rate difference of the CSOC satellites
\begin{equation}
\Delta\dot{\omega}^{\rm sun}={3a_{\odot}\over {8nae}}\left\{
\begin{array}{lllll} -2\cos i_1\cos \epsilon
\cos \Big(\Omega_1+\lambda+\omega \Big) -\\
2\Big [\Big(1-\cos i_1\Big)\cos \epsilon\Big ]
\cos \Big(\Omega_1+\lambda-\omega \Big) + \\
\Big [2\Big(1+\cos i_1\Big)\cos \epsilon\Big ]
\cos \Big(\Omega_1-\lambda+\omega \Big) + \\
-2\cos i_1\cos \epsilon
\cos \Big(\Omega_1-\lambda-\omega \Big),\\
\end{array} \right.
\label{per2}
\end{equation}
where we expressed the orbital elements of S2 in terms of those of
S1, that is $\cos i_2=-\cos i_1$ and
$\dot{\Omega}_2=-\dot{\Omega}_1$. Of course, Eq. (\ref{per2}) is
valid when the CSOC satellites are in full sun--light, i.e.,
neglecting the shadow effects due to the Earth. We can roughly
estimate the long--period effect of direct solar radiation
averaging Eq. (\ref{per2}) over our observational period $T_{\rm
obs}$. We obtain
\begin{equation}
\left\langle\Delta\dot{\omega}^{\rm sun}\right\rangle_{T_{\rm
obs}} \simeq{1 \over {T_{\rm obs}}}\int_{0}^{T_{\rm obs}}
\Delta\dot{\omega}^{\rm sun} dt\approx 150 \quad {\rm mas\
yr}^{-1}
\end{equation}
Concerning the perigee rate perturbations that arise from the
thermal thrust effects, as well those due to the Asymmetric
Reflectivity effect, they critically depend from the satellite
spin axis orientation (Lucchesi 2002a). The time evolution of the
spin axis of two satellites in supplementary inclination is quite
different. In Figures 4 and 5 the time evolution of the cartesian
components of the CSOC satellites spin axis unit--vector are
shown.
\begin{figure}[ht!]
\begin{center}
\includegraphics*[width=13cm,height=10cm]{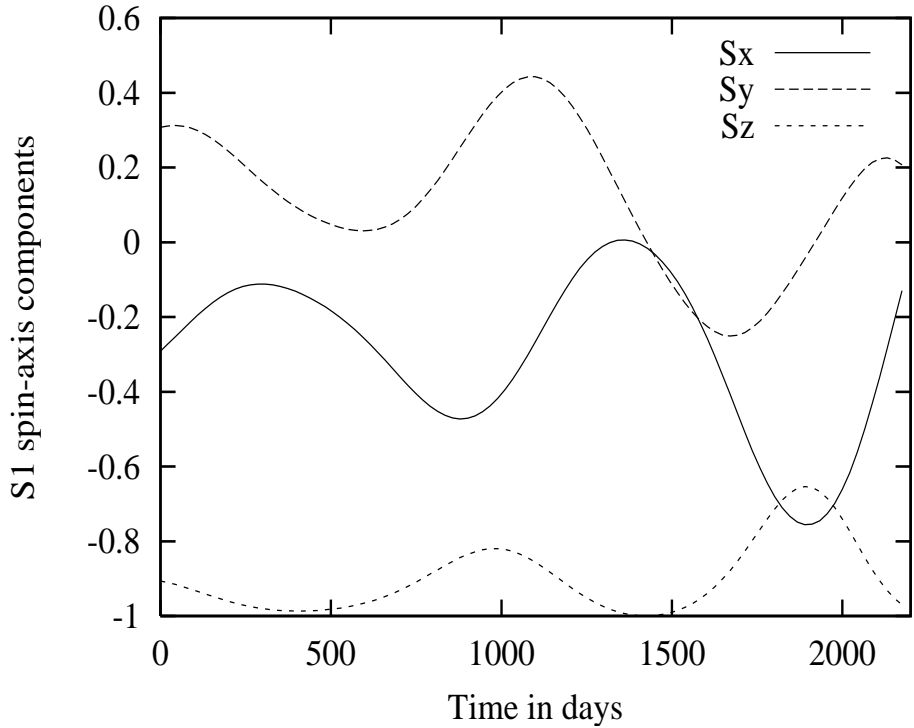}
\end{center}
\caption{\footnotesize S1 spin axis components evolution over the
IP integration time. The same magnetization parameters of LAGEOS
have been adopted.} \label{figura4}
\end{figure}
\begin{figure}[ht!]
\begin{center}
\includegraphics*[width=13cm,height=10cm]{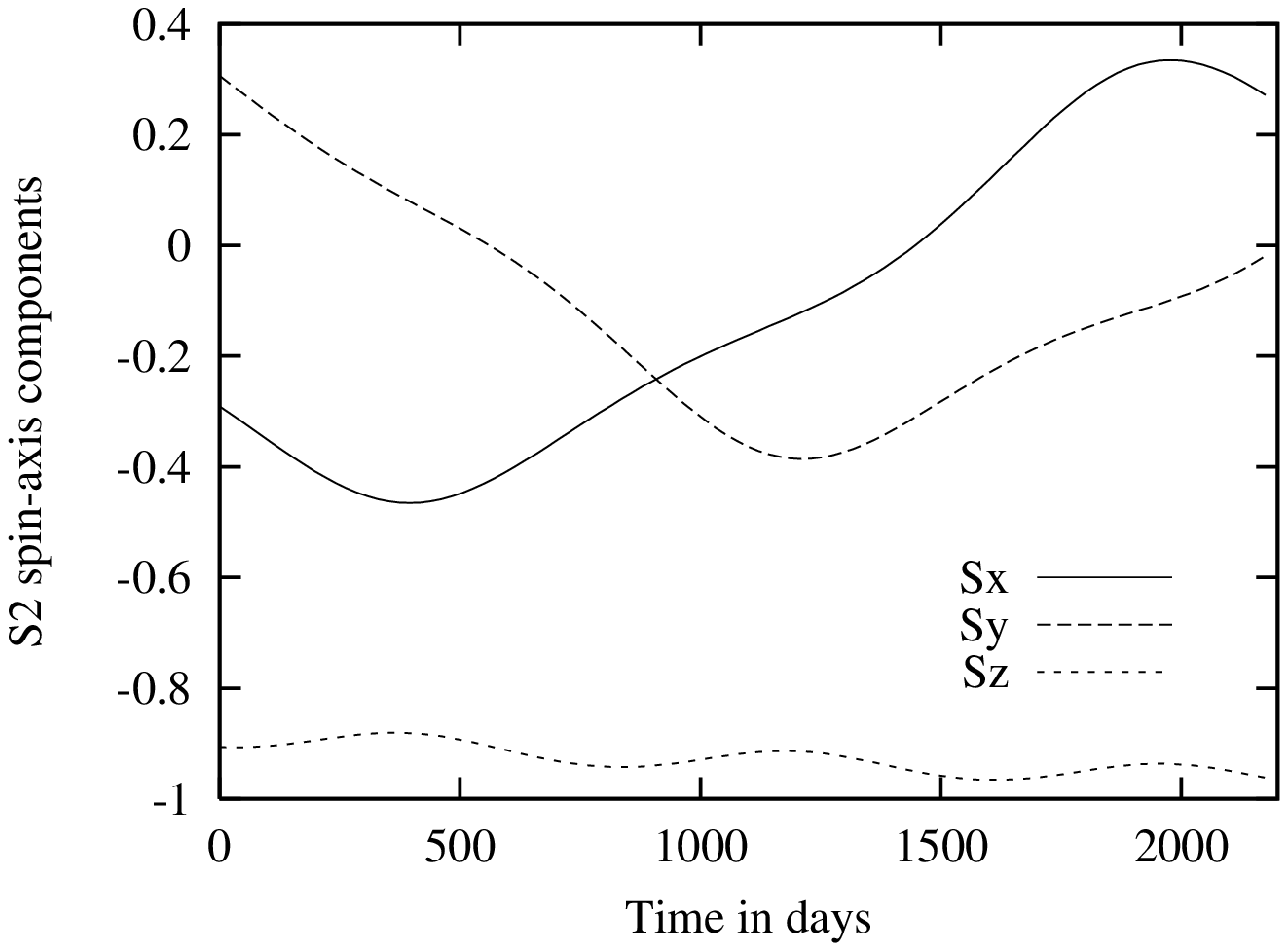}
\end{center}
\caption{\footnotesize S2 spin axis components evolution over the
IP integration time. The same magnetization parameters of LAGEOS
have been adopted.} \label{figura5}
\end{figure}
These plots have been obtained applying to our satellites the
Bertotti and Iess (1991) spin axis evolution model in the
Farinella et al. (1996) up--dated version.
%
%
%
\noindent As we can see from the plots, the equatorial components
($S_x$,$S_y$) evolution is quite different for the two satellites,
while the $S_z$ component may be considered more or less constant,
in a {\it first--approximation} approach, for both satellites.
Then, it is not so straightforwardly to compute the analytical
expressions of the Thermal Thrust effects--as well as of the
Asymmetric Reflectivity effect--for the CSOC satellites perigee
rate difference. Indeed, the expression we can compute more easily
are valid only in the case of a fixed spin axis orientation.

Of course, in the case of the terrestrial Yarkovsky--Rubincam
effect (Lucchesi 2002a), we can give the expression of the secular
rate in the perigee rate difference, because it depends only from
the $S_z$ components, that we can assume constant as previously
evidenced. We obtain
\begin{equation}
\Delta\dot{\omega}^{\rm rub}\arrowvert_{\rm sec}={A_{\rm rub}\over
{4na}}\cos \theta \Big (1-6\cos ^2 i \Big )\Big
({S_z}_1^2-{S_z}_2^2 \Big ) \label{rub}
\end{equation}
where $A_{\rm rub}$ is the amplitude of the perturbative effect
($\approx -7\times 10^{-12}$ m s$^{-2}$, assuming the same CCR
distribution of the LAGEOS satellites), while $\theta$ represents
the satellite thermal lag angle, that we have assumed to be the
same for the CSOC satellites and equal to that computed in the
case of LAGEOS (about $55^{\circ}$). Computing the averages values
of the satellites $S_z$ components over our IP we obtain, from Eq.
(\ref{rub}), a value of about -0.08 mas yr$^{-1}$, indeed a
negligible contribution from the secular effect with respect to
the relativistic $\Delta\dot{\omega}^{\rm LT}$ precession.
However, assuming a fixed spin axis, the main contributions from
the periodic terms due to the Thermal Thrust effects are the ones
we have shown in Table 2, with the addition of the harmonics
$\Omega$ and $2\Omega$ in the case of the terrestrial
Yarkovsky--Rubincam effect.
\subsection{Numerical simulation and analysis}
The orbits of the CSOC satellites have been integrated over our IP
of 2187 days with a $1^{\circ}$ step--size in the satellites
eccentric anomaly. In Tables 3 and 4 we show the results we
obtained for the analyzed NGP--in the satellites perigee and
node--neglecting any mismodelling of the perturbative effects,
i.e., their nominal amount on the elements rate.
%
\begin{table}[ht!]
\caption{NGP nominal effects on the perigee rate $\dot\omega$ (mas
yr$^{-1}$) of the proposed satellites.} \label{para}
\begin{center}
\begin{tabular}{lllllll}
\noalign{\hrule height 1.5pt} NGP & S1 & S2\\ \hline
Solar radiation & 309.86 & 212.68 \\
Earth albedo & -12.86 & -6.36 \\
Earth--Yarkovsky & 0.38 & 0.42 \\
Solar--Yarkovsky & 11.13 & 53.14 \\
Asymmetric Reflectivity & 201.28 & 206.20 \\
\noalign{\hrule height 1.5pt}
\end{tabular}
\end{center}
\end{table}
%
\begin{table}[ht!]
\caption{NGP nominal effects on the nodal rate $\dot\Omega$ (mas
yr$^{-1}$) of the proposed satellites.} \label{para}
\begin{center}
\begin{tabular}{lllllll}
\noalign{\hrule height 1.5pt} NGP & S1 & S2\\ \hline
Solar radiation & 14.77 & -20.25 \\
Earth albedo & -1.22 & 1.26 \\
Earth--Yarkovsky & -0.85 & 0.93 \\
Solar--Yarkovsky & -0.17 & $<$0.01 \\
Asymmetric Reflectivity & 0.13 & 0.11 \\
\noalign{\hrule height 1.5pt}
\end{tabular}
\end{center}
\end{table}
%
\noindent As previously pointed out, some of the analyzed NGP
effects are larger than the Lense--Thirring effect, as in the case
of the direct solar radiation and the Asymmetric Reflectivity in
the satellites perigee rate: a few hundred of mas\ yr$^{-1}$
against 44 mas\ yr$^{-1}$. In the case of the nodal rate, the
direct solar radiation perturbation gives effects comparable in
magnitude with the relativistic precession (about 33 mas\
yr$^{-1}$). Of course, the effect of the IP integration is evident
if we compare the actual results with a 2550--day  ($\approx$
7--year) integration. The average long-term effects due to direct
solar radiation become, respectively, about -778 mas yr$^{-1}$ in
the case of S1 and about 3098 mas yr$^{-1}$ in the case of S2,
i.e., more than 2.5 and 14.5 times larger! In Tables 5 and 6 we
show the results for the combination of the perigee and node
rates---with their errors with respect to the relativistic
precession---when the uncertainties of the perturbative models are
considered (the third column gives the relative uncertainty for
each model).
%
\begin{table}[ht!]
\caption{NGP effects on the CSOC--satellites perigee rate
difference $\Delta\dot\omega$ (mas yr$^{-1}$) and their errors
($\%$) with respect to the Lense--Thirring effect in the perigee
rates combination.} \label{para}
\begin{center}
\begin{tabular}{lllllll}
\noalign{\hrule height 1.5pt} NGP & $\Delta\dot\omega$ (mas yr
$^{-1}$) & Model uncertainty $\%$ & $\delta(\Delta\dot\omega)\over
\Delta\dot\omega_{\rm LT}$ $(\%)$
\\ \hline
Solar radiation & 97.18 & 0.5 & 0.55 \\
Earth albedo & -6.50 & 10 & 0.74 \\
Earth--Yarkovsky & -0.04 & 10 & $\ll 0.01$ \\
Solar--Yarkovsky & -42.01 & 10 & 4.8 \\
Asymmetric Reflectivity & -4.92 & 20 & 1.1 \\
\noalign{\hrule height 1.5pt}
\end{tabular}
\end{center}
\end{table}
%
\begin{table}[ht!]
\caption{NGP effects on the CSOC--satellites nodal rate sum
$\Sigma\dot\Omega$ (mas yr$^{-1}$) and their errors ($\%$) with
respect to the Lense--Thirring effect in the nodal rates
combination.} \label{para}
\begin{center}
\begin{tabular}{lllllll}
\noalign{\hrule height 1.5pt} NGP & $\Sigma\dot\Omega$ (mas
yr$^{-1}$) & Model uncertainty $\%$ &
${\delta(\Sigma\dot{\Omega})}\over {\Sigma\dot{\Omega}_{\rm LT}}$
$(\%)$   \\ \hline
Solar radiation & -5.47 & 0.5 & 0.04 \\
Earth albedo & 0.04 & 10 & $<$ 0.01 \\
Earth--Yarkovsky & 0.08 & 10 &  0.01 \\
Solar--Yarkovsky & -0.01 & 10 & $\ll 0.01$ \\
Asymmetric Reflectivity & 0.24 & 20 & 0.07 \\
\noalign{\hrule height 1.5pt}
\end{tabular}
\end{center}
\end{table}
%
We are now able to compare the results of the numerical analysis
with the estimates we obtained in sub--section 3.1 from the
analytic computations. For instance, the results for the direct
solar radiation pressure and the terrestrial Yarkovsky--Rubincam
effect (see Table 5) are in good agreement with those previously
obtained. Also the spectral analysis of the numerical integration
results agrees with the lines computed analytically. In figures 6
are shown the results of the Fourier analysis in the case of
direct solar radiation pressure. We have found a clear evidence of
the two strongest lines obtained with our analytical analysis,
corresponding to the spectral lines: $\Omega+\lambda\pm\omega$ and
$\Omega-\lambda\pm\omega$.
%
\begin{figure}[ht!]
\begin{center}
\includegraphics*[width=13cm,height=10cm]{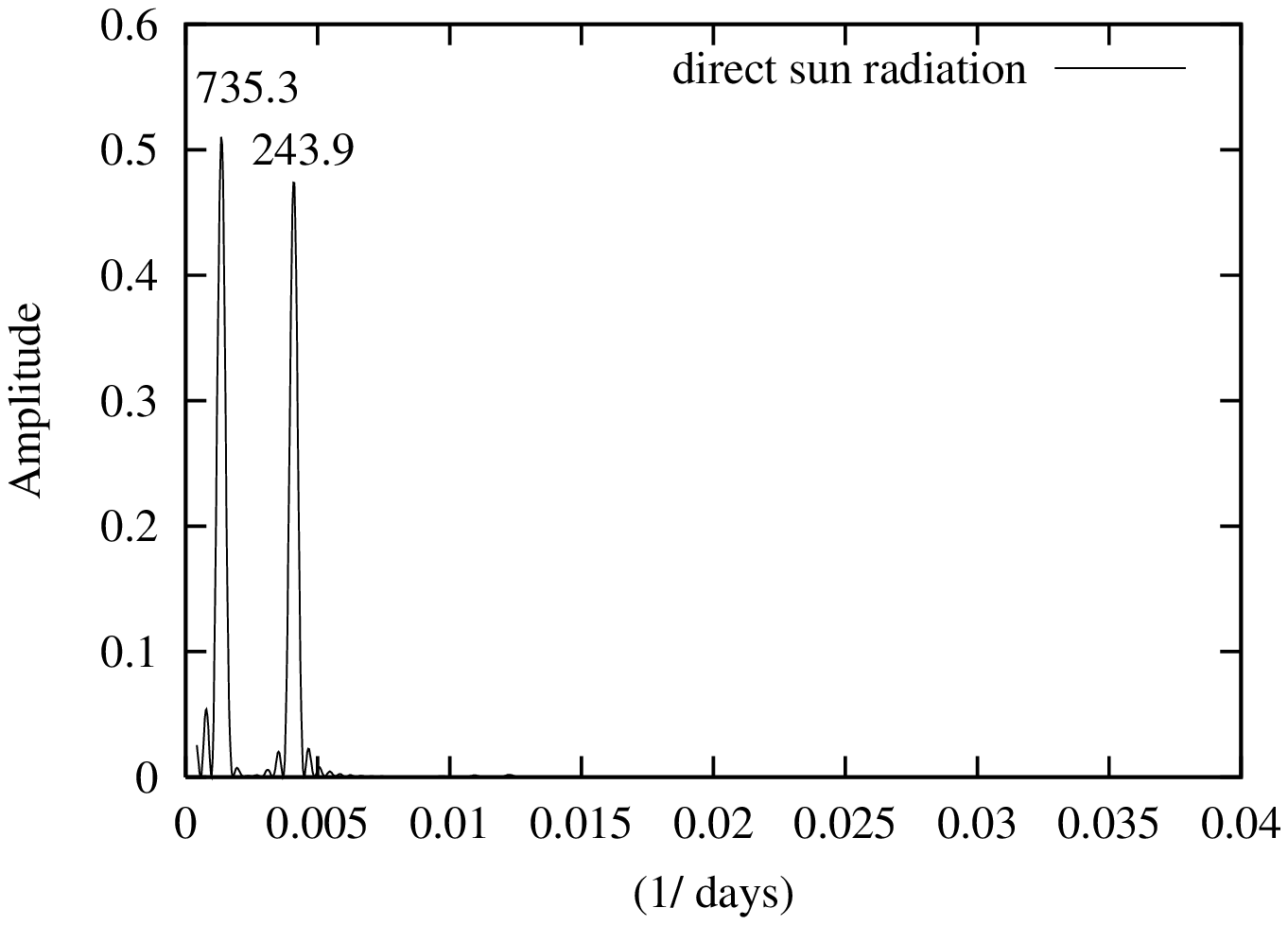}
\end{center}
\caption{\footnotesize Fourier analysis of the perigee rate
difference in the case of direct solar radiation pressure
perturbation. The vertical axis gives the normalized coefficient
of spectral correlation. } \label{figura6}
\end{figure}
%
Adding quadratically the errors for each kind of perturbation we
are able to estimate the NGP mismodelling impact on the proposed
combinations
\begin{equation}
\delta\left(\Sigma\dot{\Omega}^{\rm NGP}\right)=\left\{\sum_{\rm
pert} \left[\delta\left(\Sigma\dot{\Omega}\right)\right]^2
\right\} ^{1/2}\simeq 0.08\%\ \Sigma\dot{\Omega}^{\rm LT}
\label{nodo}
\end{equation}
\begin{equation}
\delta\left(\Delta\dot{\omega}^{\rm NGP}\right)=\left\{\sum_{\rm
pert} \left[\delta\left(\Delta\dot{\omega}\right)\right]^2
\right\}^{1/2}\simeq 5\%\ \Delta\dot{\omega}^{\rm LT}
\label{perigo}
\end{equation}
As we can see, with the sum of the nodes we obtain a smaller
impact of the NGP mismodelling on the final error budget. This is
due to the negligible influence of the Yarkovsky--Schach effect on
the node combination, as well as for the very small impact of the
direct solar radiation and the Asymmetric Reflectivity on the
proposed combination. In Table 7 we show the results for the
combination proposed with Eq. (3).
%
\begin{table}[ht!]
\caption{NGP mismodelled effects on the CSOC--satellites
nodal--rate sum and perigee--rate difference (in mas yr$^{-1}$)
and their errors ($\%$) with respect to the Lense--Thirring effect
in the $\dot X$ combination.} \label{para}
\begin{center}
\begin{tabular}{lllllll}
\noalign{\hrule height 1.5pt} NGP & $\delta(\dot{X}) $ (mas yr
$^{-1}$)& ${\delta(\dot{X})\over {\dot{X}_{\rm LT}}}$\ $(\%)$
\\ \hline
Solar radiation & 0.46 &  0.30 \\
Earth albedo & -0.65 &  0.42 \\
Earth--Yarkovsky & $<$ 0.01 & $\ll$ 0.01 \\
Solar--Yarkovsky & 4.2 &  2.72 \\
Asymmetric Reflectivity & -0.94 & 0.61 \\
\noalign{\hrule height 1.5pt}
\end{tabular}
\end{center}
\end{table}
%
With this combination the larger contribution to the NGP error
budget is due to the Yarkovsky--Schach effect. Adding again
quadratically each contribution of the NGP errors, we obtain
\begin{equation}
\delta\left(\dot{X}^{\rm NGP}\right)=\left\{\sum_{\rm pert}
\left[\delta\left(\dot{X}\right)\right]^2 \right\}^{1/2}\simeq
2.8\%\ \dot{X}^{\rm LT} \label{X}
\end{equation}
that is, a result halfway those obtained with the other
combinations. The combination introduced with Eq. (3) has the
advantage of a larger relativistic precession--about 154 mas
yr$^{-1}$--with respect to the $\Sigma\dot{\Omega}$ and
$\Delta\dot{\omega}$ combinations. In fact, this will be very
useful when computing the integrated residuals from the SLR
observations during the data-analysis. The larger slope of the
$\dot X$ combination residuals will indeed make a more clear
evidence of the total relativistic precession $\dot{X}^{\rm LT}$.
We have to stress that we have followed a quite conservative error
budget estimate, in particular concerning the uncertainty
characterizing the Asymmetric Reflectivity effect. It is also
significative to underline that actually, this very important
perturbative effect is not modelled by the orbit determination
programs used for the satellites data analyses. Nevertheless, it
is always possible to remove some of the characteristic
periodicities of this perturbation form the final fit---without
affecting the slope of the Lense--Thirring effect derivation---in
such a way to reduce the final  rms of the plotted
integrated--residuals.
%
\section{Conclusions}
In this paper we have quantitatively analyzed the scenarios
offered by the proposal of launching a pair of new twin
LAGEOS--like satellites in identical orbits and critical
supplementary inclinations (CSOC satellites) in order to measure
the gravitomagnetic Lense--Thirring effect not only by means of
the sum of their nodes but also with the difference of their
perigees. We have so intended to establish, on one hand, if the
use of the perigees would be able to yield some benefits to the
measurement of the gravitomagnetic frame dragging with respect to
the sum of the nodes, and, on the other, if the launch of a new
pair of SLR satellites would be justified also from the point of
view of the node-only observable with respect to the LAGEOS-LARES
project.

The future improvements in our knowledge of the Earth's
gravitational field thanks to the CHAMP and GRACE missions has led
us to draw our attention mainly on the impact which the
mismodelling of the non--gravitational perturbations (NGP) could
have on the proposed gravitomagnetic observables.

The sum of the nodes would yield by far the most accurate results.
The obtainable precision should be realistically considered at the
level of the order of\footnote{Here we do not consider the impact
of measurement errors like plate motion, atmosphere and polar
motion. Moreover, also the impact of the ocean  tidal
perturbations has not been addressed. In (Watkins {\it et al}
1993) six full simulations of LAGEOS--LAGEOS III data yielded a
7$\%$--8$\%$ error.} $1\%$.

The difference of the perigees would be an independent, less
accurate observable. It should be noticed that the practical data
reduction of the perigee rates should be performed very carefully
in order to account for possible, unpredictable changes in the
physical properties of the satellites' surfaces which may occur
after some years of their orbital life, as it seems it has
happened for LAGEOS II. Such effects may yield a not negligible
impact on the response to the direct solar radiation pressure.
However, the great experience obtained in dealing with the perigee
of LAGEOS II in the LAGEOS--LAGEOS II Lense--Thirring experiment
could be fully exploited  for the proposed measurement as well.
The obtainable precision for the difference in the perigees should
be of the order of $5\%$.

The combination involving the sum of the nodes with the difference
of the perigees would lie at an intermediate level of
accuracy\footnote{It should be considered that such results have
been obtained by using the force models and the approximations
which have proven to be valid for the existing LAGEOS satellites.
The new satellites could be suitably built up in order to reduce
the impact of the non-gravitational accelerations with respect to
the existing LAGEOS satellites.}.

These estimates are based on the fact that, thanks to the chosen
critical inclinations, over an observational time span of about 6
years (2187 days), all the time--dependent harmonic perturbations
would complete some full cycles. Then, they could be viewed as
empirically fitted quantities to be removed from the analyzed
temporal series: this fact should yield further improvements in
the error budget. Moreover, the impact of the orbital injection
errors on the gravitational systematic error should be probably
reduced well below $1\%$ by the new results from CHAMP and GRACE.

Finally, we must conclude that, although appealing, the use of the
alternative observable represented by the difference of the
perigees of the proposed CSOC satellites would not yield any
significant improvement with respect to the sum of the nodes as
far as the detection of the Lense-Thirring effect is concerned.
Moreover, the advantages of analyzing only the sum of the nodes of
the proposed CSOC satellites with respect to the corresponding
observable of the LAGEOS-LARES project would perhaps not justify
the expense of the construction and the launch of such entirely
new satellites, especially in view of the present-day budget
restrictions of many space agencies and of the difficulties
already encountered with the LARES.
\section*{Acknowledgements}
L. Iorio is grateful to L. Guerriero for his support while at
Bari. D. Lucchesi is grateful to P. Bonifazi and I. Ciufolini for
their support to the author research activity at IFSI/CNR (Rome).

\end{document}